# Performance Evaluation for the Co-existence of eMBB and URLLC Networks: Synchronized versus Unsynchronized TDD

Ursula Challita, Kimmo Hiltunen, and Miurel Tercero
{ursula.challita, kimmo.hiltunen, miurel.i.tercero}@ericsson.com
*Ericsson Research*

*Abstract* - **To ensure the high level of automation required in today's industrial applications, next-generation wireless networks must enable real-time control and automation of dynamic processes with the requirements of extreme low-latency and ultra-reliable communications. In this paper, we provide a performance assessment for the co-existence of a macro (eMBB) and a local factory (URLLC) network and evaluate the network conditions under which the latency and reliability requirements of factory automation applications are met. In particular, we evaluate the co-existence of the eMBB and URLLC networks under two scenarios: (i) synchronized TDD, in which both networks follow the same TDD pattern, and (ii) unsynchronized TDD, in which the eMBB and URLLC networks follow different TDD patterns. Simulation results show that the high downlink interference from the macro base stations towards the factory results in a reduction of the downlink URLLC capacity and service availability in case of synchronized TDD and a reduction of the uplink URLLC capacity and service availability in case of unsynchronized TDD. Finally, it is shown that a promising case for co-existence is the adjacent channel allocation, for both synchronized and unsynchronized TDD deployments. Here, the required isolation to protect the URLLC network in the worst-case scenario where the factory is located next to a macro site can be handled via the factory wall penetration loss (e.g., considering high concrete or metal-coated building walls) along with other solutions such as filters, larger separation distance, and band pairing.**

## I. Introduction

The fifth generation (5G) of the mobile networks is envisioned to feature different service classes: ultra-reliable low-latency communications (URLLC), massive machine type communications (mMTC), and enhanced mobile broadband (eMBB). While eMBB aims at high spectral efficiency, hard latency (e.g., 1 ms) and reliability requirements (target BLER in between $10^{-5}$ and $10^{-9}$) are essential for URLLC applications [1-4]. In essence, the stringent latency and reliability requirements of URLLC enable new emerging use cases such as factory automation, drone communication, remote control and remote surgery. An important use case for URLLC is factory automation with latency requirement of 1 ms and reliability requirement of 99.999% [1-4]. For such use case, it is crucial to assess the overall system level performance for a co-existence scenario where a local factory network has to fulfill the desired latency and reliability requirements while being interfered by the overlaid macro network offering wide area coverage in the same frequency band.

That said, an important aspect to investigate for the co-existence of a macro (eMBB) and a local (URLLC) factory network at new radio (NR) mid-band (i.e., 3.5 GHz) is the impact of utilizing different duplex patterns for the eMBB and URLLC networks, i.e., a scenario where the networks are uncoordinated. This difference rises from the fact that for the factory automation applications, the URLLC traffic is mainly symmetric in the downlink (D) and uplink (U) and thus an URLLC-optimized TDD pattern for such factory networks would be DUDU. Meanwhile, the traditional eMBB traffic is heavier in the downlink and thus a more eMBB-optimized TDD pattern for the macro network would be DDDU. Although this kind of unsynchronized TDD deployment can increase time-resource utilization, improve instantaneous data rate and decrease wireless latency, it induces new types of interference scenarios among base stations (BSs) and users (UEs) [5]. Specifically, downlink-to-uplink (BS-to-BS) and uplink-to-downlink (UE-to-UE) interference scenarios exist alongside the conventional downlink-to-downlink (BS-to-UE) and uplink-to-uplink (UE-to-BS) interference scenarios.

The co-existence between TDD macro and TDD small cells has been studied in [6]-[10]. The authors in [6] present a tutorial overview of dynamic uplink-downlink configuration in TDD systems adapting to the individual traffic needs of a specific cell area. In [7], a 3GPP technical report on dynamic TDD is presented. In particular, the benefits of uplink-downlink re-configuration as a function of traffic conditions have been evaluated. Moreover, co-existence analysis for the case of co-channel and adjacent channel interference, where adjacent channel interference may be from other operator(s), has been analyzed. In [8], the authors introduce low power almost blank subframes to alleviate the macro to small cell interference considering a dynamic TDD scenario. The authors in [9] provide co-existence analysis for two TDD networks operating over the same frequency band. Results have shown that synchronization is an essential requirement for TDD system deployment. The authors in [10] propose a scheme based on the downlink-to-uplink interference cancellation functionality in microcell BSs and/or small cell BSs in order to enable small cell dynamic TDD transmissions in heterogeneous networks. Nonetheless, the work in [6]-[10] does not consider the case of mixed traffic scenario and hence the stringent latency and reliability requirements of the small cell network are not accounted for. Moreover, none of the existing prior art studies the performance of unsynchronized TDD compared to synchronized TDD between the adjacent URLLC and eMBB networks.

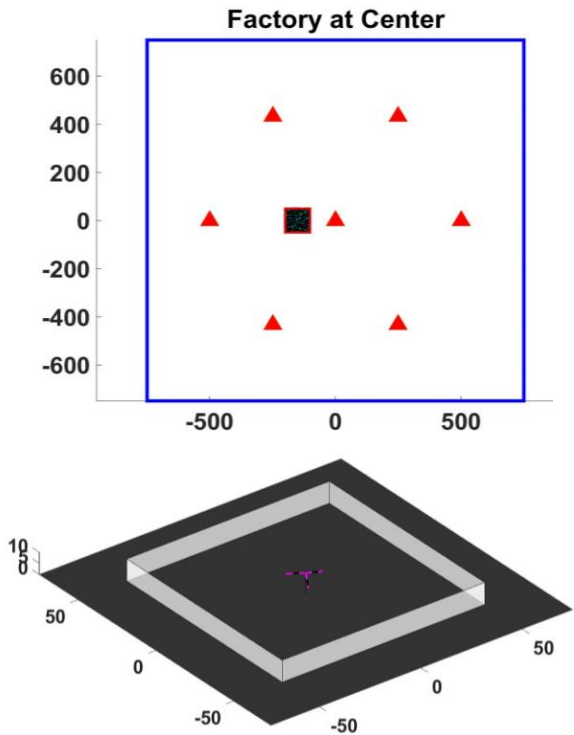

Fig. 1. Assumed network layout with seven tri-sectored macro sites (triangles) and one factory (rectangle, see also the figure below) with one tri-sectored site deployed in the middle of the factory.

The main contribution of this paper is to assess the performance of the co-existence of a macro and a local factory network under two scenarios: (i) synchronized TDD, in which both networks follow the same TDD pattern, and (ii) unsynchronized TDD, in which the macro and the local factory network follow different TDD patterns (i.e., the macro network follows an eMBB-optimized TDD pattern while the local factory network follows an URLLC-optimized TDD pattern). In particular, we consider both a co-channel and an adjacent channel deployment of the co-existing networks and provide a system-level performance analysis from both the coverage and the capacity point of view. To the best of our knowledge, this is the first work that evaluates the performance of the co-existence of eMBB and URLLC networks in the context of factory automation applications for the case of synchronized and unsynchronized TDD.

TABLE I. PROBABILITIES OF THE DIFFERENT INTER-NETWORK INTERFERENCE SCENARIOS

|  | Synchronized TDD | | Unsynchronized TDD | |
| --- | --- | --- | --- | --- |
|  | From eMBB to URLLC | From URLLC to eMBB | From eMBB to URLLC | From URLLC to eMBB |
| DL-to-DL (BS-to-UE) | 50% | 50% | 37.5% | 37.5% |
| DL-to-UL (BS-to-BS) | 0% | 0% | 37.5% | 12.5% |
| UL-to-UL (UE-to-BS) | 50% | 50% | 12.5% | 12.5% |
| UL-to-DL (UE-to-UE) | 0% | 0% | 12.5% | 37.5% |

The rest of this paper is organized as follows. In Section II, we present the system model. Simulation results and analysis are presented in Section III. Finally, conclusions are drawn in Section IV.

## II. SYSTEM MODEL

We consider an area of 1500×1500 m$^2$, as illustrated in Fig. 1, in which a macro and a local factory network are deployed. The macro network providing wide area eMBB coverage consists of seven tri-sectored sites with inter-site distance of 500 m (with wrap-around) and base station antenna height of 25 m. Meanwhile, for the local factory network offering URLLC connectivity, we consider a single factory of 100×100×10 m$^3$ with one tri-sectored ceiling-mounted site deployed in the middle of the factory, pointing horizontally with a specific down-tilt. We assume that the URLLC users are uniformly distributed inside the factory, while all eMBB users are located outdoors and no eMBB users are located inside the factory. Moreover, we consider three different factory locations, thus realizing the different impact from/to the macro network: cell-edge, center, and near-BS.

We assume that the macro and the factory networks are operating in the 3.5 GHz frequency band and apply TDD as the duplexing method. Two different TDD deployments are evaluated:

- Unsynchronized TDD: The macro network follows a DDDU TDD pattern while the local factory network follows a DUDU TDD pattern.

- Synchronized TDD: Both networks follow a DUDU TDD pattern.

The slot borders are assumed to be aligned for both synchronized and unsynchronized TDD configuration. Finally, the resulting probabilities for the different inter-network interference scenarios are given in Table I. Here, it is important to note that the considered TDD patterns are chosen as an example for comparison purposes of synchronized and unsynchronized TDD. Another reasonable TDD pattern is to consider an eMBB-optimized DDDU pattern for both networks for the case of synchronized TDD.

## A. Propagation Model

We assume the 3GPP Urban Macro propagation model [11] for the links between the macro base stations and the eMBB users, and the 3GPP Indoor Hotspot Open Office model [11] for the links between the factory base stations and the indoor URLLC users. Furthermore, the path losses between the macro base stations and the users or base stations inside the factory are calculated as a combination of the 3GPP Urban Macro propagation model, wall penetration loss and an indoor loss. Meanwhile, the path losses between the factory base stations or users and the outdoor eMBB users are calculated as a combination of the 3GPP Urban Micro propagation model [11], wall penetration loss and an indoor loss. The wall penetration loss is modeled as a function of the wall material and frequency band and it accounts for the angular loss that is a function of the incident angle [12]. In this study, we assume that the wall penetration loss (for perpendicular penetration) is equal to 13 dB, corresponding to an average loss for a wall consisting approximately of 93% concrete and 7% traditional two-pane windows [11]. Furthermore, the simulation results are compared against "full isolation" in which case the wall loss has been assumed to be equal to infinity. Finally, the indoor loss is expressed as $D \cdot d_{in}$ where $D$ is 0.5 dB/m as in [11] and $d_{in}$ is the travelled indoor distance.

## B. Performance Metrics

The URLLC users are assumed to be successfully served if they can fulfill the reliability requirement of 99.999% within a latency bound of 1 ms. In practice, the desired QoS cannot be guaranteed if a) the maximum achievable user bit rate is less than what would be required to transmit the message payload during one TTI, or b) the system does not have enough radio resources to successfully serve the total network offered load. For the performance evaluation, we consider the following URLLC metrics:

- URLLC service availability: Percentage of locations within the factory floor where the desired QoS can be guaranteed. We consider a uniform sampling across the factory floor where $i$ corresponds to a particular sample and N is the total number of samples. The URLLC service availability, $SA_{URLLC}$, can be expressed as:

$$SA_{URLLC} = \left(\frac{\sum_{i=1}^{N} x_i}{N}\right) \times 100$$

  where $x_i=1$ if the desired QoS can be guaranteed and $x_i=0$ otherwise.
- URLLC system capacity: Maximum packet arrival rate at which the 100% URLLC service availability can still be reached. Service availability equal to 100%

TABLE II. SIMULATION PARAMETERS

| Parameter | Factory Network | Macro Network |
|---|---|---|
| Radio access technology | NR | NR |
| Frequency [GHz] | 3.5 | 3.5 |
| Bandwidth [MHz] | 50 | 50 |
| Duplex: Synchronized TDD Unsynchronized TDD | DUDU DUDU | DUDU DDDU |
| DL:UL traffic ratio | 1:1 | 1:1 |
| Sectors per site | 3 | 3 |
| BS transmit power [dBm] | 27 | 50 |
| UE transmit power [dBm] | 23 | 23 |
| UE Antenna Gain [dBi] (isotropic) | 0 | 0 |
| BS noise figure [dB] | 5 | 5 |
| UE noise figure [dB] | 9 | 9 |
| Max BS antenna element gain [dBi] | 8 | 8 |
| BS antenna array V x H x (Vs x Hs x Ps) | 2x4x(2x1x2) | 8x8x(1x1x2) |
| SNR-based uplink power control | Alpha=0.8 target SNR=10dB | Alpha=0.8 target SNR=10dB |

is essential for factory applications to guarantee continuous service throughout the factory floor.

For performance assessment, we consider both a co-channel and an adjacent channel deployment. First, the impact of the inter-network interference on the coverage i.e., URLLC service availability and the average eMBB bit rates is evaluated for a co-channel deployment, assuming a packet size of 32 Bytes and a fixed level of offered area traffic for both networks (5 packets/s/m$^2$ for URLLC and 100 Mbps/km$^2$ (low eMBB) or 300 Mbps/km$^2$ (high eMBB) for eMBB). Second, the impact of the inter-network interference on the URLLC system capacity is evaluated for an adjacent channel deployment.

## III. SIMULATION RESULTS AND ANALYSIS

In this section, we summarize the main findings for the assumed co-existence scenario between a local URLLC factory network and an overlaid macro eMBB network. The main focus of the study is on the impact of the eMBB network interference on the performance of the factory network. However, the impact on the performance of the eMBB network is also briefly discussed.

For performance evaluation, a simulator is used where the eMBB and URLLC networks are modeled with some statistical model and considering different traffic models and arrival rates. Table II provides a summary of the main simulation parameters for both networks considering NR mid-band at 3.5 GHz. For the URLLC network, a subcarrier spacing of 30 kHz and packet size of 32 Bytes are assumed. A transmission time interval (TTI) length of 143 µs is considered with 4 OFDM symbols per TTI. Moreover, we consider QPSK, 16 QAM, and 64 QAM for the available modulation and coding schemes of the URLLC network with the corresponding (1/20, 1/10, 1/5, 1/3), (1/3, 1/2, 2/3), and (2/3, 3/4) code rates, respectively. Next, we summarize the main findings considering both a co-channel and an adjacent channel deployment.

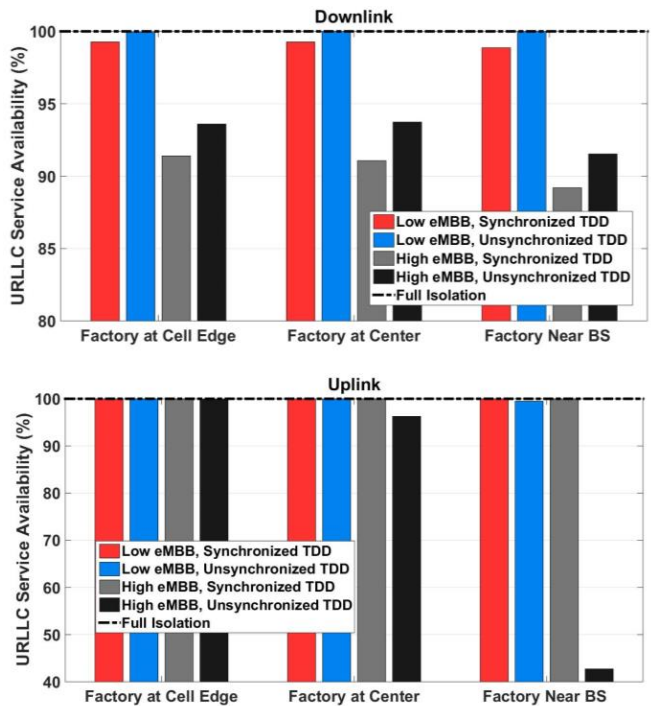

Fig. 2. Downlink and uplink URLLC service availability for the different factory locations.

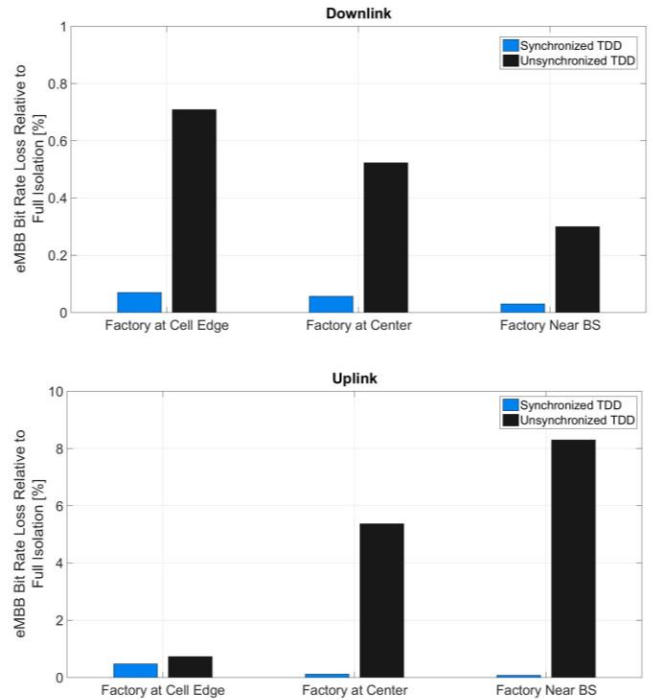

Fig. 3. Downlink and uplink eMBB performance losses for the different factory locations.

### A. Co-channel Deployment

For a co-channel deployment, the local factory and the macro networks are assumed to be sharing the same channel, and the main objective is to evaluate both the URLLC service availability inside the factory and the average eMBB bit rates outside the factory. Fig. 2 presents the results for downlink and uplink URLLC service availability for the different factory locations with respect to the macro site. Assuming an unsynchronized TDD deployment, full URLLC service availability can be achieved in the downlink for all factory locations if a low eMBB load is assumed (corresponding to an average macro cell utilization of approximately 20%), while with a high level of inter-network interference (average macro cell utilization of approximately 90%) the URLLC service availability drops to 92-94%. If a synchronized TDD deployment is assumed instead, the downlink URLLC service availability becomes clearly worse, and the full URLLC service availability cannot be observed for any of the factory locations, not even with the low level of eMBB load. There are two main reasons why the synchronized TDD results in a worse downlink URLLC performance compared to the unsynchronized TDD: a) the URLLC downlink is constantly interfered by high-power macro base stations, and b) the average downlink cell utilization of the macro network is increased from 20% to 30% (low eMBB) or from 90% to 100% (high eMBB) as a result of the change in the TDD pattern from DDDU to DUDU. Therefore, the level of the inter-network interference towards the URLLC downlink is increased at the time instances of downlink transmissions, resulting in worse downlink SINR values, and consequently in a worse downlink URLLC service availability as some of the users will not be able to reach their minimum required downlink bit rates.

Meanwhile, the situation looks the opposite for the uplink URLLC service availability. In case of unsynchronized TDD, the factory base stations are part of the time interfered by the downlink transmissions from the high-power macro base stations (cross-link interference between the base stations), which can have a very large negative impact on the uplink URLLC service availability in particular when the factory is located close to the macro site and if the load in the macro network is high (100% resource utilization in this case). However, if the networks are synchronized, full uplink URLLC service availability can be secured for all three factory locations. Again, there are two main reasons why synchronized TDD is so beneficial for the URLLC uplink performance in this case: a) factory base stations are interfered only by the power-controlled eMBB users located outside the factory, b) the amount of uplink time domain resources is doubled for the eMBB users resulting in considerably reduced average macro cell utilizations (reduced from 100% to 60% in case of low eMBB). As a result, the level of inter-network interference experienced by the factory base stations becomes considerably lower, improving the uplink SINR values, and finally improving the URLLC service availability since more users can reach their minimum required uplink bit rates.

TABLE III. AVERAGE eMBB BIT RATE (GAIN IN DOWNLINK AND LOSS IN UPLINK) FOR UNSYNCHRONIZED TDD COMPARED TO SYNCHRONIZED TDD FOR THE CASE OF FULL ISOLATION.

|  | Factory Location | | |
|---|---|---|---|
|  | **Cell Edge** | **Center** | **Near BS** |
| **Downlink** | +76.8 % | +66.2 % | +55.4 % |
| **Uplink** | -53.7 % | -57.3 % | -54.4 % |

When it comes to the impact of the inter-network interference towards the macro network, we consider the scenario with a low eMBB load (100 Mbps/km$^2$). The impact on downlink performance is evaluated by looking at the average bit rate of the eMBB users within a 15 m polygon surrounding the factory, while the impact on the uplink performance is evaluated by looking at the average bit rate of the closest macro sector. Fig. 3 shows the observed eMBB performance loss for synchronized and unsynchronized TDD compared to full isolation. As can be seen, the impact of the inter-network interference on the eMBB users is in general small in the downlink. The downlink performance losses are higher when the networks are unsynchronized, which can be explained by a lower level of the intra-network interference (due to a lower level of average cell utilization) resulting in a higher impact of the inter-network interference. Furthermore, the performance losses are the higher, the further away from the serving macro base station the victim users are located. However, even though the impact of the inter-network interference is higher for unsynchronized TDD, the overall eMBB downlink performance is still better due to the larger amount of time domain resources compared to the synchronized TDD. This can be clearly seen from Table III which summarizes the difference between the average eMBB bit rate for unsynchronized TDD relative to synchronized TDD for the case of full isolation, i.e, when the impact of the inter-network interference is ignored.

Looking at the uplink results for unsynchronized TDD, it can be noticed that the impact of the inter-network interference is clearly higher compared to the downlink. This is caused by the cross-link interference from the factory base stations towards the macro base stations. Furthermore, another disadvantage of the unsynchronized TDD is that the amount of time domain uplink resources is halved compared to the synchronized TDD, which results in clearly worse average eMBB bit rates even when a full isolation between the networks is assumed, as demonstrated by the values in Table III.

### B. Adjacent Channel Deployment

For an adjacent channel deployment, we study the required level of isolation between the networks so that the maximum URLLC system capacity is not affected by the inter-network interference. Here, we assume a fully-loaded macro network. Results for the downlink and uplink URLLC system capacity with respect to a scenario with full isolation between the networks are shown in Fig. 4. As can be noticed, a slightly lower level of isolation is required in the downlink for the case of unsynchronized TDD compared to synchronized TDD. It becomes also clear that the highest level of additional isolation, approximately 60 dB, is required when the factory is located next to the macro site. In uplink, however, a much higher level of isolation is required for the unsynchronized TDD compared to synchronized TDD. In case of synchronized TDD, the assumed wall penetration loss of 13 dB is sufficient to protect the URLLC network from any capacity losses, while in case of unsynchronized TDD an additional isolation of 55 dB is required when the factory is located next to the macro site.

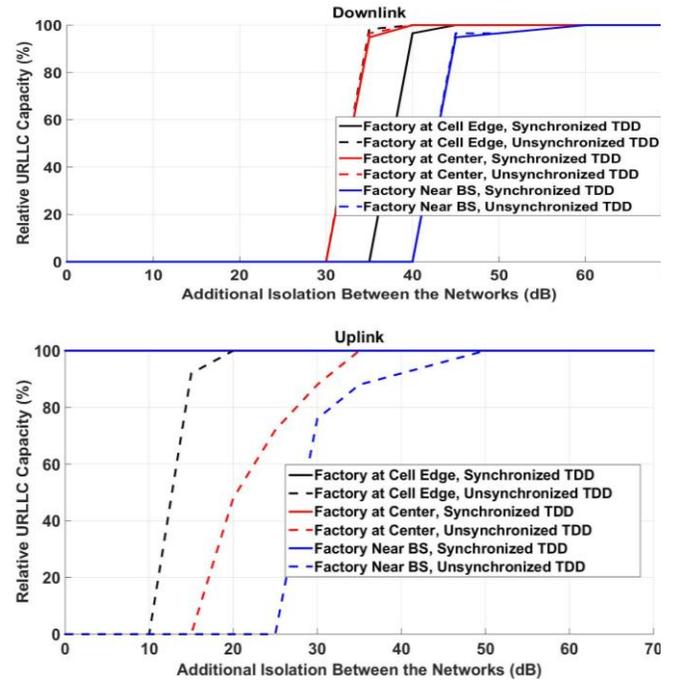

Fig. 4. Relative downlink and uplink URLLC system capacity as a function of the additional isolation between the networks on top of the assumed wall loss of 13 dB.

In case of an adjacent channel deployment between the two networks, part of the required isolation is offered by the adjacent channel leakage ratio (ACLR) and the adjacent channel selectivity (ACS) of the involved transmitters and the receivers, respectively. In case of the synchronized TDD, the overall adjacent channel interference ratio (ACIR) would be limited to approximately 30 dB for both the downlink and the uplink due the UE characteristics (assuming for simplicity that the ACLR and ACS values are equal to 45 dB for the BS and 30 dB for the UE [13]). This means that the remaining 30 dB of the required isolation between the networks should be taken care of by an additional wall penetration loss or some other means. In case of unsynchronized TDD, a separate ACIR value would be applied for each inter-network interference scenario: 30 dB for downlink-to-downlink (BS-to-UE) and uplink-to-uplink (UE-to-BS), 42 dB for downlink-to-uplink (BS-to-BS) and 27 dB for uplink-to-downlink (UE-to-UE). In general, this means that in case of downlink where the required isolation is in the order of 60 dB for the worst-case deployment to cope with the high level of interference from the macro base stations towards the URLLC users, approximately 30 dB can be taken care of by the ACIR, while the remaining 30 dB have to be taken care of by other means. In case of unsynchronized TDD in uplink, the problems are related to the very high level of cross-link interference from the macro base stations towards the factory base stations. Here, most of the required isolation of 55 dB can be taken care of by the ACIR (42 dB), while the remaining 13 dB must be taken care of by some other means, such as increased wall penetration loss, factory site densification, and uplink power control.

However, it is also worth highlighting that the results presented here assume already a concrete wall with fairly small window areas. For a solid concrete wall, or assuming that the traditional windows would be replaced by modern energy-efficient windows, the wall loss would increase to approximately 19 dB [11], i.e., proving an additional isolation of 6 dB compared to the results shown above. Hence, in order to be able to protect the URLLC system capacity even within the worst-case deployment, some other means to either reduce the level of the inter-network interference, or to reduce the impact of the inter-network interference are required. As an example, the level of the inter-network interference can be lowered for example with metal-coated building walls, by avoiding deploying high-power macro sites close to the factory building, or by pointing the close-by macro base station antennas away from the factory. Furthermore, the impact of the inter-network interference can be reduced by densifying the factory network, or by increasing the transmission power of the factory base stations and the URLLC UEs.

## IV. CONCLUSION

In this paper, we have evaluated the performance of a co-existence scenario between an eMBB macro network and a local URLLC factory network with different network load levels as well as with different TDD patterns for both networks. Results have shown that the high downlink interference from the macro base stations towards the factory results in a reduction of the downlink URLLC capacity and service availability in case of synchronized TDD and a reduction of the uplink URLLC capacity and service availability in case of unsynchronized TDD. Furthermore, the results confirm that a promising case for co-existence is the adjacent channel allocation, for both synchronized and unsynchronized TDD deployments. A local factory URLLC network can co-exist with an eMBB network when an isolation of approximately 73 dB is guaranteed to protect the URLLC network in the worst-case scenario where the factory is located next to a macro site. Here, most of the required isolation can be taken care of by the adjacent channel attenuation (42 dB), while the remaining isolation can be handled by some other means, such as increased wall penetration loss (considering metal-coated or thick concrete building walls), factory site densification, uplink power control, larger separation distance, and band pairing.

As part of future work, it is important to investigate and evaluate interference coordination mechanisms both in time and frequency domain (i.e., coordinated scheduling avoiding the most harmful collisions between the neighboring networks) and in power domain (i.e., controlling the base station and UE powers so that the interference between the networks can be limited to reduce both the level and the impact of the inter-network interference). Moreover, it is crucial to assess a co-existence scenario with a denser factory network, as well as a scenario with adjacent channel eMBB users located inside the factory. Another important future direction is the co-existence scenarios within the mmWave bands. It is also interesting to evaluate the impact of other TDD patterns for the considered scenario. Here, note that one can qualitatively estimate the system performance under different TDD patterns. For instance, in case the DDDU pattern is considered for both networks for synchronized TDD, the URLLC network would not be capable of supporting the tight latency requirement of 1 ms in the uplink and would not be efficient for balanced uplink/downlink traffic volumes. For the actual performance numbers with different TDD patterns, separate set of system level evaluations would be needed.